\documentclass[11pt,a4paper]{article}

\usepackage{amssymb}
\usepackage{amsmath}
\usepackage{epsfig}
\usepackage{amsthm}

\newtheorem{patheorem}{Theorem}[section]
\newtheorem{paproposition}[patheorem]{Proposition}
\newtheorem{padefinition}[patheorem]{Definition}
\newtheorem{paexample}[patheorem]{Example}
\newtheorem{palemma}[patheorem]{Lemma}
\newtheorem{pacorollary}[patheorem]{Corollary}

\newenvironment{theorem}{\begin{patheorem}\rm }{\end{patheorem}}

\newenvironment{example}{\begin{paexample}\rm }{{\hfill \proved}\end{paexample}}
\newenvironment{lemma}{\begin{palemma}\rm }{\end{palemma}}

\newcommand{\proved}{\hfill $\Box$}

\newcommand{\tpntuple}{\tuple{\places,\transitions,\inputs,\outputs}}
\newcommand{\tpn}{N}
\newcommand{\places}{P}
\newcommand{\place}{p}
\newcommand{\qplace}{q}
\newcommand{\transitions}{T}
\newcommand{\transition}{t}
\newcommand{\timedtrans}{\longrightarrow_{\it Time}}
\newcommand{\disctrans}{\longrightarrow_{\it Disc}}
\newcommand{\ttrans}{\longrightarrow_{\transition}}
\newcommand{\inputs}{{\it In}}
\newcommand{\outputs}{{\it Out}}
\newcommand{\Intervals}{{\it Intrv}}
\newcommand{\interval}{{\cal I}}
\newcommand{\jinterval}{{\cal J}}
\newcommand\intrvl[2]{\left[{#1},{#2}\right]}
\newcommand{\marking}{M}
\newcommand{\initmarking}{\marking_{\it init}}
\newcommand{\markings}{{\sf M}}
\newcommand{\markingset}{{\sf M}}
\newcommand{\finalmarkings}{{\sf M}_{\it fin}}

\newcommand{\trans}{\longrightarrow}
\newcommand{\transs}{\stackrel{*}{\longrightarrow}}

\newcommand{\constr}{\phi}
\newcommand{\constrs}{\Phi}

\newcommand{\constrf}{\constr_F}
\newcommand{\conf}{\gamma}
\newcommand{\confs}{\Gamma}
\newcommand{\confinit}{\conf_{{\it init}}}
\newcommand{\conff}{\conf_F}
\newcommand{\set}[1]{\left\{ #1 \right\}}
\newcommand{\tuple}[1]{\left( #1 \right)}
\newcommand{\setcomp}[2]{\left\{#1|\;#2\right\}}

\newcommand{\entailed}{\preceq}
\newcommand{\confentailed}{\preceq_{\confs}}

\newcommand{\pre}{{\it Pre}}
\newcommand{\pred}{{\pre_{\it Disc}}}
\newcommand{\pret}{\pre_{\transition}}
\newcommand{\preT}{\pre_{\it Time}}

\newcommand{\nat}{{\cal N}}
\newcommand{\preals}{{\cal R}^{\geq 0}}
\newcommand{\upto}[1]{\underline{#1}^1}
\newcommand{\uptoz}[1]{\underline{#1}^0}

\newcommand{\integers}{{\cal Z}}

\newcommand{\zone}{Z}
\newcommand{\region}{R}
\newcommand{\denotationof}[1]{[\![#1]\!]}
\newcommand{\normal}[1]{\widetilde{#1}}
\newcommand{\conjunct}[3]{{#1}\otimes\tuple{{#2},{#3}}}
\newcommand{\add}[3]{{#1}\oplus\tuple{{#2},{#3}}}
\newcommand{\abstr}[2]{{#1}\backslash{#2}}
\newcommand{\covered}{\sqsubseteq}

\newcommand{\funtype}[3]{{#1}:{#2}\rightarrow{#3}}
\newcommand{\vertices}{{\tt V}}
\newcommand{\vertex}{{\tt v}}
\newcommand{\edges}{{\tt E}}
\newcommand{\refvertex}{{\tt v}^0}
\newcommand{\reffvertex}{{\tt v}^1}
\newcommand{\diff}{D}

\newcommand{\powerset}{{\cal P}}

\newcommand{\var}{v}

\newcommand{\uvar}{u}

\newcommand{\seq}{s}
\newcommand{\seqsf}{\nat^{<\omega}}
\newcommand{\seqsi}{\nat^{\omega}}
\newcommand{\symbols}{\lambda}
\newcommand{\tail}{{\it tail}}
\newcommand{\barrier}{\beta}
\newcommand{\abarrier}{\alpha}
\newcommand{\str}{w}
\newcommand{\ddd}{Y}
\newcommand{\positive}{{\it pos}}
\newcommand{\negative}{{\it neg}}
\newcommand{\op}{{\it op}}
\newcommand{\low}{{\it low}}
\newcommand{\high}{{\it high}}
\newcommand{\const}{{\it const}}

\newcommand{\natset}[1]{\widehat{#1}}

\newcommand{\constrsystem}{{\mathbb C}}

\newcommand{\sparseconstrs}{{\mathbb S}_1}
\newcommand{\exsparseconstrs}{{\mathbb S}_2}
\newcommand{\lcsconstrs}{{\mathbb L}}
\newcommand{\broadconstrs}{{\mathbb B}}

\newcommand{\mstar}[1]{{#1}^\oplus}
\newcommand{\mset}{B}
\newcommand{\msetleq}{\leq}

\newcommand{\placing}{\bar{P}}
\newcommand{\cmax}{{\it cmax}}
\newcommand{\zones}{{\sf Z}}
\newcommand{\regions}{{\sf R}}

\newcommand{\wpreceq}{\preceq_w}

\begin{document}
\author{Parosh Aziz Abdulla \and Aletta Nyl\'en \and {\small\begin{tabular}{c} Dept. of Information Technology, Uppsala University\\ P.O. Box 337, SE-751 05 Uppsala, Sweden\\Email:\{parosh, aletta\}@it.uu.se\end{tabular}}}

\title{Better Quasi-Ordered Transition Systems\thanks{
Parts of this paper have appeared 
in Proc. LICS'2000, 14th IEEE Int. Symp. on Logic in Computer Science,
and 
Proc. ICATPN'2001, 22nd Int. Conf. on application and theory of Petri nets.}}

\date{}

\maketitle
\begin{abstract}
Many existing algorithms for model checking of infinite-state systems
operate on {\em constraints} which are used to represent (potentially
infinite) sets of states. A general powerful technique which can be
employed for proving termination of these algorithms is that of {\em
  well quasi-orderings}. Several methodologies have been proposed for
derivation of new well quasi-ordered constraint systems. However, many
of these constraint systems suffer from a ``constraint explosion
problem'', as the number of the generated constraints grows
exponentially with the size of the problem. In this paper, we
demonstrate that a refinement of the theory of well quasi-orderings,
called the theory of {\em better quasi-orderings}, is more appropriate
for symbolic model checking, since it allows inventing constraint
systems which are both well quasi-ordered and compact. As a main
application, we introduce {\it existential zones}, a constraint system
for verification of systems with unboundedly many clocks and use our
methodology to prove that existential zones are better
quasi-ordered. We show how to use existential zones in verification of
timed Petri nets and present some experimental results. Also, we apply
our methodology to derive new constraint systems for verification of
broadcast protocols, lossy channel systems, and integral relational
automata. The new constraint systems are exponentially more succinct
than existing ones, and their well quasi-ordering  cannot be shown by
previous methods in the literature.
\end{abstract}

\section{Introduction}
A major current challenge in automatic program verification is to
extend model checking \cite{CES:modelchecking,QuSi:cesar} to
transition systems with infinite state spaces. Standard techniques
such as reachability analysis and tableau procedures can be adapted,
by using {\em constraints} to represent (potentially infinite) sets of
states. These algorithms are based on two operations, namely that of
computing predecessors or successors of sets of states (represented by
constraints), and that of checking for termination (formulated as
entailment between constraints). Since the number of constraints is
not {\it a priori} bounded, a key problem when applying the
algorithms, is to guarantee termination. A general powerful tool which
can be applied for proving termination is to show that the set of
constraints is {\em well quasi-ordered} under entailment. In
\cite{Parosh:Bengt:Karlis:Tsay:general,Parosh:Bengt:tutorial}, a
constraint based backward reachability algorithm is
presented. Furthermore, a methodology is defined for inventing well
quasi-ordered constraint systems. The key idea is to start from a set
of ``basic'' constraints, and repeatedly derive new ones, using the
fact that well quasi-orderings are closed under certain operations on
constraints such as building finite trees, words, vectors, multisets,
sets, etc. The methodology has been applied both to unify earlier
existing results for Petri nets, timed automata, lossy channel
systems, completely specified protocols, relational automata, etc, and
to design verification algorithms for new classes of systems such as
timed networks \cite{Parosh:Bengt:Timed:Networks}, broadcast protocols
\cite{Esparza:Finkel:Mayr:LICS,Delzanno:etal:broadcast:constr}, and
cache coherence protocols \cite{Delzanno:cache}. However many of the
constraint systems constructed according to this method, suffer from a
``constraint explosion'' problem, as the number of constraints
generated when computing predecessors (successors) grows exponentially
with the number of components.

In this work, we demonstrate that a refinement of the theory
of well quasi-orderings, called the theory of {\em better quasi-orderings}
\cite{Milner:BQO,Pouzet:BQO} is more appropriate for symbolic model
checking, as it allows for constraint systems which are more compact
and hence less prone to constraint explosion. More precisely,
better quasi-orderings offer two advantages:
(i) better quasi-ordering implies well quasi-ordering; hence all the 
verification algorithms originally designed for
well quasi-ordered constraint systems are also 
applicable to better quasi-ordered ones;
and
(ii) better quasi-orderings are more ``robust'' than well quasi-orderings.
For instance, in addition to the above mentioned operations,
better quasi-ordered constraint systems (in contrast to
well quasi-ordered ones) are closed under disjunction:
if a set of constraints is better quasi-ordered under
entailment, then the set of finite disjunctions of these constraints
is also better quasi-ordered under entailment.
In this paper, we provide several examples which show that using disjunction
often leads to very compact  constraint systems.

First, we propose a new constraint system, which we call
{\em existential zones}, for verification of systems
with unboundedly many clocks such as timed networks
\cite{Parosh:Bengt:Timed:Networks} and timed Petri nets
(Section~\ref{tpn:section}).
Such systems cannot be modelled as real-time automata,
since the latter operate on a finite set of clocks.
An existential zone specifies a minimal required behaviour,
typically of the form $\exists x_1 x_2:\;3 \leq x_2-x_1 \leq 8$,
characterizing the set of configurations
in which there exist {\em at least} two clocks whose values differ by 
at least $3$ and at most $8$. Existential zones are related to {\em
  existential regions},  which are used in
\cite{Parosh:Bengt:Timed:Networks} for verification of timed
networks. Existential regions are better quasi-ordered since they are
constructed by repeatedly building words, multisets, and sets. Each
existential zone is equivalent to the disjunction of a finite number
of existential regions. Since better quasi-orderings are closed under
disjunction, it follows that existential zones are better quasi-ordered
(and hence well quasi-ordered).
The well quasi-ordering of existential zones cannot be shown
using the approach of 
\cite{Parosh:Bengt:Karlis:Tsay:general,Parosh:Bengt:tutorial,Finkel:Schnoebelen:everywhere},
since well quasi-orderings in general are not closed under disjunction.
In fact, an existential zone is often equivalent to the disjunction
of an exponential number of existensial regions, thus offering a much
more compact representation
(in the same manner that
{\em zones} are more efficient than {\em regions}
in verification tools for real-time automata
\cite{Uppaal:nutshell,Yovine:kronos}).
We can extend the results further and consider ``existential variants'' of CDDs
\cite{CDD:basic} and DDDs \cite{DDD:basic}-- constraint systems which 
are even more compact than zones.
We have implemented a prototype based on existential DDDs, and carried out
a verification of a parametrized version of Fischer's protocol.
While the set of constraints explodes when using existential regions, our tool
performs the verification in a few seconds.

We also consider broadcast protocols, which consist of an arbitrary number
of finite-state processes, communicating through rendezvous or through
broadcast.
In \cite{Esparza:Finkel:Mayr:LICS} safety properties are checked, 
using constraints which characterize upward closed sets
of vectors of natural numbers.
In \cite{Delzanno:etal:broadcast:constr} several new constraint
systems are proposed, represented by different forms
of linear inequalities over natural numbers.
Since the new constraint systems cannot be constructed from
upward closed sets using the previously mentioned constraint operations,
these classes require an explicit termination proof for the
underlying reachability algorithm.
Applying our methodology we are able to prove well
quasi-ordering of these constraint systems in a uniform manner.
More precisely, the upward closed sets are characterized by constraints
which are vectors of natural numbers and hence are better quasi-ordered.
Using the fact that each inequality is a finite union (disjunction) of
upward closed sets, we conclude that they are also better
quasi-ordered (and hence well quasi-ordered).

Finally, we provide new better quasi-ordered constraint systems for
verification of lossy channel systems \cite{AbJo:lossy:IC} and 
integral relational automata \cite{Cerans:relational:automata:ICALP}.
The new constraint systems are exponentially more succinct than existing ones.

\paragraph{Related Work}
The first work which applies well quasi-orderings in
symbolic model checking is reported in \cite{AbJo:lossy}.
The main contribution was an algorithm for checking
safety  properties for {\it lossy channel systems}.
The idea  of the algorithm is to perform  {\it backward} reachability
analysis using the fact that the underlying transition relation is 
{\it monotonic} under a given well quasi-ordering.

Independently, Finkel \cite{Finkel:completely:specified} 
used well quasi-orderings for checking termination properties.
This algorithm uses forward analysis and is therefore not sufficiently
powerful for verification of safety properties.

The method of \cite{AbJo:lossy} was extended in
\cite{Parosh:Bengt:Karlis:Tsay:general} into a general framework for
verification of relational automata, (Timed) Petri nets, timed
networks, etc. In \cite{Finkel:Schnoebelen:everywhere} the
monotonicity conditions of  \cite{Parosh:Bengt:Karlis:Tsay:general}
were further relaxed extending applicability to new classes of
systems.

To our knowledge this work is the first application of the theory of
better quasi-orderings in the context of symbolic model checking.

Existential zones are variants of zones, a symbolic representation used
in several tools for verification of timed automata, e.g. KRONOS
\cite{Yovine:kronos} and UPPAAL \cite{Uppaal:nutshell}. However, zones
characterize finite sets of clocks and therefore cannot be used to
analyze timed Petri nets.

A model close to timed Petri nets, timed networks, was considered in
\cite{Parosh:Bengt:Timed:Networks:journal}. A timed network consists
of an arbitrary number of timed processes and hence contain an
unbounded number of clocks. The constraint system used in that work
was that of existential regions, a constraint system that is far less
efficient than existential zones and the number of existential regions
generated during analysis explode even on small applications.

Most earlier work on studying decidability issues for timed Petri
nets, e.g. \cite{Razouk:Phelps:TPN,Berthomieu:Diaz:TPN,GMMP:TPN,PCVC99:TPN:Nondecidability}, 
either report undecidability results or
decidability under the assumption that the net is bounded. A work
closely related to ours is \cite{Escrig:etal:TPN}. The authors
consider the coverability problem for a class of timed Petri nets
similar to our model. The main difference is that in
\cite{Escrig:etal:TPN}, it is assumed that the ages of the tokens are
natural numbers. Furthermore, it is not evident how efficient the
constraint system is in practical applications.

\paragraph{Outline}
In the next Section we introduce the notions of constraints
and well quasi-orderings.
In Section~\ref{bqo:basics:section} and
Section~\ref{bqo:application:section} we give the basics 
of the theory of better quasi-orderings and its application in model checking.
Timed Petri nets are defined in
Section~\ref{tpn:section} and in Section~\ref{zones:section} we
introduce existential zones and show how they can be used in the
analysis of timed Petri nets. The results of our experiments are
described in Section~\ref{implementation:section}. In
Section~\ref{broadcast:section}, Section~\ref{lcs:section} and
Section~\ref{ira:section} we provide better quasi-ordered constraint
systems for the verification of broadcast protocols, lossy channel
systems and integral relational automata, respectively.
Finally in Section~\ref{conclusion:section} 
we give some conclusions and directions for future work. 

\section{Constraints and WQOs}
\label{overview:section}

In this section, we introduce the notions of {\em constraints}
and {\em well quasi-orderings},
and describe how to use them for performing symbolic model checking.
We assume a transition system $\tuple{\confs,\trans}$,
where $\confs$ is a potentially infinite set of 
{\em configurations}, and
$\trans$ is a transition relation on $\confs$ whose reflexive transitive
closure is denoted by $\transs$.

\paragraph*{Constraints}
We use {\em constraints} $\phi$ for representing sets 
$\denotationof{\constr}$ of configurations.
We define an {\em entailment relation $\preceq$} on constraints,
where $\constr_1\entailed\constr_2$ iff
$\denotationof{\constr_2}\subseteq \denotationof{\constr_1}$, and
let $\equiv$ be the equivalence relation induced by $\entailed$.
We sometimes write disjunctions $\constr_1\vee\cdots\vee\constr_n$
of constraints as $\vee\set{\constr_1,\ldots,\constr_n}$.
For sets $\constrs_1,\constrs_2$ of constraints, we let
$\constrs_1\covered\constrs_2$ denote that for each
$\constr_2\in\constrs_2$ there is a $\constr_1\in\constrs_1$ with
$\constr_1\entailed\constr_2$.
Notice that $\constrs_1\covered\constrs_2$ implies 
$\vee\constrs_1\entailed\vee\constrs_2$.

\paragraph*{Reachability}
In the sequel, we concentrate on the {\em reachability problem}:
given a configuration
$\confinit$ and a constraint $\constrf$, is there
$\conff\in\denotationof{\constrf}$ such that $\confinit\transs\conff$?
We perform a backward reachability analysis, generating a sequence 
$\constrs_0\subseteq\constrs_1\subseteq\constrs_2\subseteq\cdots$
of finite sets of constraints 
where $\constrs_0=\set{\constrf}$ and
$\constrs_{j+1}=\constrs_j\cup\pre(\constrs_j)$.
Here $\pre(\constrs)=\cup_{\constr\in\constrs}\pre(\constr)$, where
$\pre(\constr)$ is a finite set of constraints, such that
$\denotationof{\vee\pre(\constr)}=
\setcomp{\conf'}{\exists\conf\in\denotationof{\constr}.\; \conf'\trans\conf}$.
For all the constraint systems we consider in this paper, the
set $\pre(\constr)$ exists and is computable.
Since $\constrs_0\sqsupseteq\constrs_1\sqsupseteq\constrs_2\sqsupseteq\cdots$,
the algorithm terminates when we reach a point $j$ where 
$\constrs_j\covered\constrs_{j+1}$
(implying $\vee\constrs_{j+1}\equiv\vee\constrs_j$).
Then, $\constrs_j$ 
characterizes the set of all predecessors of $\constrf$
(sometimes written  as $\pre^*(\constrf)$).
This means that the answer to the reachability question is equivalent
to whether $\confinit\in\denotationof{\vee\constrs_j}$. We observe that,
in order to be able to implement the algorithm for a given class of systems,
the constraint system should allow
(i)
computing $\pre(\constr)$,
(ii)
checking entailment between constraints, and
satisfiability of a constraint by a configuration.

To show termination we rely on the theory of {\em well quasi-orderings (wqo)}.
A constraint system is said to be {\em well quasi-ordered}
if for each infinite sequence 
$\constr_0,\constr_1,\constr_2,\ldots$ of constraints, there are
$i<j$ with $\constr_i\entailed\constr_j$.
The following lemma 
(from \cite{Parosh:Bengt:Karlis:Tsay:general})
characterizes the class of constraint systems for which termination
is guaranteed.
\begin{lemma}
\label{termination:lemma}
A constraint system is well quasi-ordered iff for each
infinite ($\subseteq$-increasing) sequence
$\constrs_0\subseteq \constrs_1\subseteq \constrs_2\subseteq\cdots$ 
of constraint sets,
there is a $j$ such that $\constrs_j \sqsubseteq \constrs_{j+1}$.
\end{lemma}

\paragraph*{Remark on Well Quasi-Ordered Transition Systems}
Alternatively, we can consider transition systems $\tuple{\confs,\trans}$
which are  {\it well quasi-ordered} \cite{Parosh:Bengt:Karlis:Tsay:general,Parosh:Bengt:tutorial,Finkel:Schnoebelen:everywhere}.
This means that the set $\confs$ of configurations is equipped with 
a well quasi-ordering $\confentailed$ such that the transition 
relation is {\it monotonic} with respect to $\confentailed$.
In other words, for configurations
$\conf_1,\conf'_1,\conf_2$, if
$\conf_1\confentailed\conf'_1$ and 
$\conf_1\trans\conf_2$ then there is a configuration
$\conf'_2$ such that $\conf_2\confentailed\conf'_2$ and
$\conf'_1\trans\conf'_2$.
We can now develop a theory based on well quasi-ordered 
transitions systems rather than well quasi-ordered constraint systems.
The two theories are intimately related and yield identical 
model checking algorithms
\cite{Parosh:Bengt:Karlis:Tsay:general,Parosh:Bengt:tutorial}.
All constraints which we we will consider in this paper
characterize sets of configurations which are upward closed with respect
to $\confentailed$.
This means that the reachability problem described above in fact asks
about reachability of sets of states which are upward closed sets 
of states rather than that of a single state.
This offers two advantages:
\begin{itemize}
\item
Checking safety properties amounts to upward closed set reachability.
More precisely,
the states in $\denotationof\conff$ are usually taken to be bad states that
we do not want to occur during an execution. Using standard techniques
\cite{GoWo:safety,VW:modelchecking}, we can reduce several classes of
safety properties to the reachability problem.
\item
Single state reachability is more difficult to solve.
For instance, in the context of Petri nets, upward closed set
reachability amounts to {\it coverability}.
In {\it timed Petri nets}, single state reachability is undecidable
\cite{PCVC99:TPN:Nondecidability}, while we show in this paper that
coverability  is decidable.
\end{itemize}

\section{Basics of BQOs}
\label{bqo:basics:section}
In this section, we introduce the basic definitions and properties of better
quasi-orderings.
We let $\nat$ denote the set of natural numbers,
and let $\seqsf$ ($\seqsi$) denote the set of finite (infinite) 
strictly increasing sequences over $\nat$.
For $\seq\in\seqsf$, we let $\symbols(\seq)$ be the set of natural numbers
occurring in $\seq$, and if $\seq$ is not empty then we let
$\tail(\seq)$ be the result of deleting the first element of $\seq$.
For $\seq_1\in\seqsf$ and $\seq_2\in\seqsf\cup\seqsi$, we 
write $\seq_1\ll\seq_2$ to denote that $\seq_1$ is a proper prefix of 
$\seq_2$.
If $\seq_1$ is not empty then we write $\seq_1\ll_*\seq_2$ to denote 
that $\tail(\seq_1)\ll\seq_2$.
An infinite set $\barrier\subseteq\seqsf$ is said to be a barrier if the following 
two conditions are satisfied.
\begin{itemize}
\item
There are no $\seq_1,\seq_2\in\barrier$ such that 
$\symbols(\seq_1)\subsetneq\symbols(\seq_2)$.
\item
For each $\seq_2\in\seqsi$ there is $\seq_1\in\barrier$ with
$\seq_1\ll\seq_2$.
\end{itemize}

Let $\tuple{A,\preceq}$ be a quasi-ordering, i.e.,
$\preceq$ is a reflexive and transitive
relation on $A$.
An {\em $A$-pattern} is a mapping $\funtype{f}{\barrier}{A}$, where
$\barrier$ is a barrier.
We say that $f$ is {\em good} if there are $\seq_1,\seq_2\in\barrier$ such
that $\seq_1\ll_*\seq_2$ and $f(\seq_1)\preceq f(\seq_2)$.
We say that $\tuple{A,\preceq}$ is a {\em better quasi-ordering (bqo)}
if each $A$-pattern is good.

We use $A^{\omega}$ to denote the set of infinite sequences over $A$.
For $\str\in A^{\omega}$, we let $\str(i)$ be the the $i^{th}$
element of $\str$.
For a quasi-ordering $\tuple{A,\preceq}$, we define
the quasi-ordering $\tuple{A^{\omega},\preceq^{\omega}}$ where
$\str_1\preceq^{\omega}\str_2$ if and only if there is a
strictly monotone\footnote{
meaning that $h(j_1)< h(j_2)$ if and only if $j_1< j_2$}
injection $\funtype{h}{\nat}{\nat}$
such that $\str_1(i)\preceq\str_2(h(i))$, for each $i\in\nat$.

We shall use the following two properties (from \cite{Milner:BQO})

\begin{lemma}
\label{bqo:basic:lemma}
\ \\[-2em]\begin{itemize}
\item
If $\barrier$ is a barrier and $\barrier=\barrier_1\cup\barrier_2$, then
there is a barrier $\abarrier$ such that $\abarrier\subseteq\barrier_1$ or
$\abarrier\subseteq\barrier_2$.
(using induction on $n$ we can generalize this property to
$\barrier=\barrier_1\cup\cdots\cup\barrier_n$).
\item
If $\tuple{A,\preceq}$ is bqo then $\tuple{A^{\omega},\preceq^{\omega}}$ is bqo
\end{itemize}
\end{lemma}

\section{Application of BQOs}
\label{bqo:application:section}
As evident from Lemma~\ref{termination:lemma},
well quasi-ordering is crucial for
termination of the symbolic algorithm presented in 
Section~\ref{overview:section}.
Furthermore, three other properties of a given constraint system decide how 
efficient the algorithm may run in practice.
These properties are the size of the set $\pre(\constr)$,
the cost of checking entailment and membership, and the number
of iterations needed before achieving termination.
In
\cite{Parosh:Bengt:Karlis:Tsay:general,Parosh:Bengt:tutorial,Finkel:Schnoebelen:everywhere}, 
a methodology is defined for inventing well quasi-ordered constraint systems,
based on the fact that all finite domains are well quasi-ordered under equality,
and that well quasi-orderings are closed
under a basic set of operations including building finite trees, words, vectors, multisets,
sets, etc.
This means that we can start from a set of constraints over finite domains, and then
repeatedly generate new constraints by building more compound data structures.
A typical application of this approach is a constraint system,
called {\em existential regions}, introduced in \cite{Parosh:Bengt:Timed:Networks}
for verification of systems with unboundedly many clocks.
However, constraints developed according to the above methodology
suffer from ``constraint explosion'' caused by the size of the
set $\pre(\constr)$.
For instance, using existential regions, the set of generated
constraints explodes even for very small examples.
Often, the constraint explosion can be much reduced, by
considering new constraint systems,
which are disjunctions of the ones derived using the above mentioned set
of operations.
In Section~\ref{zones:section} we present {\em existential zones}
each of which corresponds to the disjunction of a (sometimes exponential)
number of existential regions.
Thus, existential zones offer a much more compact representation,
allowing us to verify a parameterized 
version of Fischer's protocol in a few seconds.
As we show in this section, well quasi-ordered constraint systems
are not closed under disjunction, and hence we cannot prove 
well quasi-ordering of existential zones within the framework
of 
\cite{Parosh:Bengt:Karlis:Tsay:general,Parosh:Bengt:tutorial,Finkel:Schnoebelen:everywhere}.

Instead of wqos, we propose here to use an alternative approach based on
bqos.
In Theorem~\ref{bqo:properties:theorem}, we state
some properties of bqos which make them attractive for
symbolic model checking.
In the rest of this section we write 
$\tuple{A,\preceq}$ to denote a quasi-ordering $\preceq$ on a set
$A$.
Let $A^*$ denote the set of finite words over $A$, and 
let $\mstar{A}$ denote the set of finite multisets over $A$.
For a natural number $n$, let $\natset{n}$ denote the set
$\{1,\ldots,n\}$.
An element $w$ of $A^*$ and of $\mstar{A}$ can be represented as
a mapping $w\colon\natset{|w|}\mapsto A$
where $|w|$ is the size of the multiset or the length of the sequence.
Given a quasi-order $\preceq$ on a set $A$,
define the quasi-order $\preceq^*$ on $A^*$ by letting
$w\preceq^* w'$ if and only if there is a strictly monotone
injection
$h\colon\natset{|w|}\mapsto\natset{|w'|}$ such that
$w(j)\preceq w'(h(j))$ for $1 \leq j \leq |w|$.
Define the quasi-order $\mstar{\preceq}$ on $\mstar{A}$ by
$w\mstar{\preceq} w'$ if and only if there is a 
(not necessarily monotone) injection
$h\colon\natset{|w|}\mapsto\natset{|w'|}$ such that
$w(j)\preceq w'(h(j))$ for $1 \leq j \leq |w|$.

In the following theorem we state some properties of bqos which we 
use later in the paper.
The proof of property~\ref{bqo:disj:prop} is in  \cite{Marcone:entail:bqo}.
We use  $\powerset(A)$ to denote the powerset of $A$.
\begin{theorem}
\label{bqo:properties:theorem}
\ \\[-2em]\begin{enumerate}
\item
\label{bqo:relation:prop}
Each bqo is wqo.
\item
\label{finite:bqo:prop}
If $A$ is finite, then $\tuple{A,=}$ is bqo.
\item
\label{bqo:str:prop}
If $(A,\preceq)$ is bqo, then $(A^*,\preceq^*)$ is bqo.
\item
\label{bqo:multiset:prop}
If $(A,\preceq)$ is bqo, then $(\mstar{A},\mstar{\preceq})$ is bqo.
\item
\label{bqo:disj:prop}
If $\tuple{A,\preceq}$ is bqo, then $\tuple{\powerset(A),\covered}$
is bqo \cite{Marcone:entail:bqo}\footnote{\cite{Jancar:csd}
provides a proof for a weaker version of the theorem, 
namely that bqo of $\tuple{A,\preceq}$ 
is sufficient for wqo of $\tuple{\powerset(A),\covered}$.}.
\end{enumerate}
\end{theorem}
\begin{proof}
We show properties \ref{bqo:relation:prop}-\ref{bqo:multiset:prop}.

\ref{bqo:relation:prop}. Follows immediately from definitions of bqo and wqo.

\ref{finite:bqo:prop}. 
Consider $\tuple{A,=}$ where $A=\set{a_1,\ldots,a_n}$ is finite.
Let $\funtype{f}{\barrier}{A}$ be an $A$-pattern.
Define $\barrier_i=f^{-1}(a_i)$, for $i:1\leq i\leq  n$.
By Lemma~\ref{bqo:basic:lemma}, there is a barrier $\abarrier\subseteq\barrier_i$,
for some $i:1\leq i\leq n$.
Take any $\seq_1\in\abarrier$ and any $\seq_2\in\seqsi$, where
$\seq_1\ll_*\seq_2$.
Since $\abarrier$ is a barrier, we know that there is $\seq_3\in\abarrier$
such that $\seq_3\ll\seq_2$, and that $\seq_3\not\subseteq\seq_1$.
It follows that $\seq_1\ll_*\seq_3$ and hence $f$ is  good.

\ref{bqo:str:prop}.
Suppose that $\tuple{A,\preceq}$ is bqo.
We show that $\tuple{A^*,\preceq^*}$ is bqo.
Take any $b\not\in A$.
For $\str\in A^*$, we let $\str'$ denote $\str b^{\omega}$
(i.e., we add infinitely many $b$:s to the end of $\str$).
It is clear that $\str_1\preceq^* \str_2$ if and only if
$\str'_1\preceq^{\omega} \str'_2$.
Let $\funtype{f}{\barrier}{A^*}$ be an $A^*$-pattern.
We know that $\funtype{f'}{\barrier}{A^{\omega}}$,
where $f'(\seq)=\str'$ iff $f(\seq)=\str$, is an $A^{\omega}$-pattern.
By Lemma~\ref{bqo:basic:lemma} it follows that there are
$\seq_1,\seq_2\in\barrier$ such that $\seq_1\ll_*\seq_2$ and
$f'(\seq_1)\preceq^{\omega}f'(\seq_2)$, and hence
$f(\seq_1)\preceq^*f(\seq_2)$.
This means that $f$ is good.

\ref{bqo:multiset:prop}. Follows from \ref{bqo:str:prop}.
\end{proof}

\begin{figure*}
\begin{center}
\includegraphics{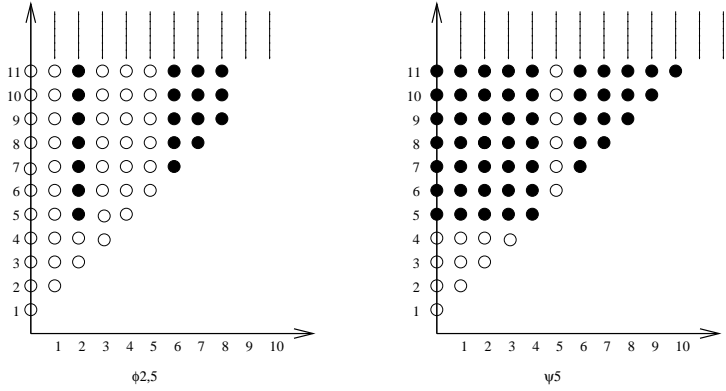}
\caption{A graphic illustration of $\denotationof{\phi_{2,5}}$ and
$\denotationof{\psi_5}$. Filled circles represent points satisfying
the corresponding constraint.}
\label{phipsi:fig}
\end{center}
\end{figure*}

Since bqo is a stronger relation than wqo (property \ref{bqo:relation:prop}),
it follows by Lemma~\ref{termination:lemma} that,
to prove termination of the reachability algorithm of
Section~\ref{overview:section}, it is sufficient to prove bqo
of constraints under entailment.
All constraint systems derived 
earlier in the literature based on the approach of
\cite{Parosh:Bengt:Karlis:Tsay:general,Parosh:Bengt:tutorial,Finkel:Schnoebelen:everywhere}) use properties \ref{finite:bqo:prop},
\ref{bqo:str:prop}, and \ref{bqo:multiset:prop}.
This implies that all these constraint systems are also bqos.
An immediate consequence of property~\ref{bqo:disj:prop} is that
bqo of a set of constraints implies bqo of disjunctions of these constraints.

In the next sections, we introduce several constraint systems applying
the following two steps.
\begin{enumerate}
\item
We show better quasi-ordering of a constraint system $\constrsystem_1$
using properties \ref{finite:bqo:prop},
\ref{bqo:str:prop}, and \ref{bqo:multiset:prop} in
Theorem~\ref{bqo:properties:theorem} (following a similar methodology to that
described in 
\cite{Parosh:Bengt:Karlis:Tsay:general,Parosh:Bengt:tutorial,Finkel:Schnoebelen:everywhere}).
\item
We use property~\ref{bqo:disj:prop} to
derive better quasi-ordering of a new more compact 
constraint system $\constrsystem_2$ defined as disjunctions of constraints in
$\constrsystem_1$.
\end{enumerate}
We notice that, although $\constrsystem_2$ is more compact, the computational complexity
for checking membership and entailment  may be higher for 
$\constrsystem_2$ than for $\constrsystem_1$.
Furthermore, the reachability algorithm of 
Section~\ref{overview:section} needs in general a higher
number of iterations in case $\constrsystem_2$ is employed.
However, in almost all cases, the compactness offered by
$\constrsystem_2$ is the dominating factor in the efficiency 
of the algorithm.

As mentioned earlier, an important difference compared to
the approach of \cite{Parosh:Bengt:Karlis:Tsay:general,Parosh:Bengt:tutorial,Finkel:Schnoebelen:everywhere}) 
is that Step 2 (taking disjunction)
cannot be performed within that framework.
This is illustrated by the following example, which shows
that wqos in general are not closed under disjunction.

\begin{example}{\bf [Rado's Example]}
\label{Rado:example}
Consider the the set $X=\setcomp{\tuple{a,b}}{a<b}\subseteq\nat^2$.
Define a set $\constrsystem_1=\setcomp{\constr_{a,b}}{\tuple{a,b}\in X}$
of constraints, such that 
the denotation $\denotationof{\constr_{a,b}}\subseteq X$ of $\constr_{a,b}$
is the set
$\setcomp{\tuple{c,d}}{(c>b) \vee ((c=a)\wedge(d\geq b))}$.
It is straightforward to check that $\constrsystem_1$ is wqo:
suppose that we have sequence 
$\constr_{a_1,b_1},\constr_{a_2,b_2},\ldots$, where
$\constr_{a_i,b_i}\not\entailed\constr_{a_j,b_j}$ if $i<j$.
Consider the sequence
$\tuple{a_1,b_1},\tuple{a_2,b_2},\ldots$.
First, we show that $a_j\leq b_1$ for all $j\geq 1$.
If this is not the case, then let $b_1 < a_j$.
We show that this implies $\constr_{a_1,b_1}\entailed\constr_{a_j,b_j}$
which is a contradiction.
Take any $(c,d)\in\denotationof{\constr_{a_j,b_j}}$.
Then, either $c>b_j$ or $(c=a_j)\wedge(d\geq b_j)$.
In both cases, we show that $c>a_1$ and hence $(c,d)\in\denotationof{\constr_{a_1,b_1}}$.
\begin{itemize}
\item
$c>b_j$.
We have $b_j>a_j$ and $b_1>a_1$ by definition, 
and $a_j>b_1$ by assumption.
It follows that $c>a_1$.
\item
$(c=a_j)\wedge(d\geq b_j)$.
We know that $b_1>a_1$ by definition, and
$a_j>b_1$ by assumption.
It follows that $c>a_1$.
\end{itemize}
Since $a_j\leq b_1$ for all $j\geq 1$,
we have a subsequence of the form
$\tuple{a,b_{i_1}},\tuple{a,b_{i_2}},\ldots$, and hence there are
$k$ and $\ell$ such that $b_{i_k} \leq b_{i_{\ell}}$ which is a contradiction.

Now, we consider a set 
$\constrsystem_2$ of constraints of the form
$\psi_j$, where 
$\psi_j\equiv\phi_{0,j}\vee\cdots\vee\phi_{j-1,j}$.
The sequence $\psi_1,\psi_2,\ldots$ violates the wqo property,
since for each $k,\ell:k<\ell$, we have 
$\tuple{k,\ell}\in\denotationof{\psi_{\ell}}$,
but $\tuple{k,\ell}\not\in\denotationof{\psi_k}$, and hence
$\denotationof{\psi_{\ell}}\not\subseteq\denotationof{\psi_k}$.
In Figure~\ref{phipsi:fig}, we give
graphic illustrations of $\denotationof{\phi_{2,5}}$ and
$\denotationof{\psi_5}$.

\end{example}
\section{Timed Petri Nets}
\label{tpn:section}
We consider {\em Timed Petri Nets (TPNs)} where each token is equipped
with a real-valued clock representing the ``age'' of the token.
The firing conditions of a transition include the usual ones for Petri
nets. Furthermore, each arc between a place and a transition is labeled
with a subinterval of the natural numbers.
When a transition is fired, the tokens removed from the input places
of the transition and the tokens added to the output places should
have ages lying in the intervals of the corresponding arcs.

We let $\integers$ and $\preals$ denote the sets of 
integers, and nonnegative reals respectively.
Recall that $\nat$ denotes the set of natural numbers.

We also recall that $\mstar{A}$ denotes the set of finite 
multisets over $A$.
Often, we view a multiset $\mset$ over a set $A$ as
a mapping from $A$ to $\nat$.
Sometimes we write multisets as lists, so e.g.\
$\tuple{2.4,5.1,5.1,2.4,2.4}$ represents a multiset $\mset$ over $\preals$
where $\mset(2.4)=3$, $\mset(5.1)=2$ and $\mset(x)=0$ for $x\neq
2.4,5.1$.
We may also write $\mset$ as $\tuple{2.4^3,5.1^2}$.
For multisets $\mset_1$ and $\mset_2$ over a set $A$, we say that
$\mset_1\msetleq\mset_2$ if $\mset_1(a)\leq\mset_2(a)$ for each $a\in A$.
We define $\mset_1+\mset_2$ to be the multiset $\mset$ where
$\mset(a)=\mset_1(a)+\mset_2(a)$, and (assuming $\mset_1\msetleq\mset_2$) we
define $\mset_2-\mset_1$ to be the multiset $\mset$ where
$\mset(a)=\mset_2(a)-\mset_1(a)$, for each $a\in A$.
We use $\emptyset$ to denote the empty multiset, i.e., $\emptyset(a)=0$ for
each $a\in A$.

We use a set $\Intervals$ of {\em intervals}. An open interval is 
written as $(w,z)$ where $w\in\nat$ and $z\in\nat\cup\set{\infty}$. 
Intervals can also be closed in one or both directions, e.g. $[w,z)$
is closed to the left. For $x\in\preals$, we write $x\in\intrvl a b$
to denote that $a\leq x\leq b$.

A {\em Timed Petri Net (TPN)} is a tuple $\tpn=\tpntuple$ where
$\places$ is a finite set of {\em places}, $\transitions$ is a finite
set of {\em transitions} and
$\inputs,\outputs:\transitions\times\places\mapsto\mstar{\Intervals}$.
If $\inputs(\transition,\place)(\interval)\neq\emptyset$ 
($\outputs(\transition,\place)(\interval)\neq\emptyset$), for
some interval $\interval$, we say that $\place$ is an 
{\em input (output) place} of $\transition$.

\paragraph{Markings}
A {\em marking $\marking$} of $\tpn$ is a finite multiset over
$\places\times\preals$.
The marking $\marking$ defines numbers and ages
of the tokens in each place in the net.
That is, $\marking(\place,x)$ defines the number of tokens with age
$x$ in place $\place$.
For example, if 
$\marking=\tuple{\tuple{\place_1,2.5},\tuple{\place_1,1.3},
\tuple{\place_2,4.7},\tuple{\place_2,4.7}}$, then,
in the marking $\marking$, there are two tokens  with ages
$2.5$ and $1.3$ in $\place_1$, 
and two tokens each with age $4.7$ in the place $\place_2$. 
Abusing notation, we define, for each place $\place$, a multiset
$\marking(\place)$ over $\preals$, where
$\marking(\place)(x)=\marking(p,x)$. Notice that untimed Petri nets
are a special case in our model where all intervals are of the form
$[0, \infty)$.

\paragraph*{Transition Relation}
We define two types of transition relations on markings.
A {\em timed transition} increases the age of all tokens
by the same real number.
Formally $\marking_1\timedtrans\marking_2$ 
if $\marking_1$ is of the form
$\tuple{\tuple{\place_1,x_1},\ldots,\tuple{\place_n,x_n}}$, and there
is $\delta\in\preals$ such that 
$\marking_2=\tuple{\tuple{\place_1,x_1+\delta},\ldots,\tuple{\place_n,x_n+\delta}}$.

We define the set of {\em discrete transitions $\disctrans$} as
$\bigcup_{\transition\in\transitions}\ttrans$, where
$\ttrans$ represents the effect of firing the transition $\transition$.
More precisely, we define $\marking_1\ttrans\marking_2$ if, for each
place $\place$ with
$\inputs(\transition,\place)=\tuple{\interval_1,\ldots,\interval_m}$
and
$\outputs(\transition,\place)=\tuple{\jinterval_1,\ldots,\jinterval_n}$,
there are multisets $\mset_1=\tuple{x_1,\ldots,x_m}$ and
$\mset_2=\tuple{y_1,\ldots,y_n}$ over $\preals$, such that the
following holds.
\begin{itemize}
\item
$\mset_1\leq\marking_1(\place)$.
\item
$x_i\in\interval_i$, for $i:1\leq i\leq m$.
\item
$y_i\in\jinterval_i$, for $i:1\leq i\leq n$.
\item
$\marking_2(p)=\left(\marking_1(p)-\mset_1\right)+\mset_2$.
\end{itemize}
Intuitively, a transition $\transition$ may be fired only if for each
incoming arc to the transition, there is a token with the ``right'' age
in the corresponding input place.
This token will be removed from the input place when the transition
is fired.
Furthermore, for each outgoing arc, a token with an age in the interval
will be added to the output place. 
We define the relation
$\trans$ to be $\timedtrans\cup\disctrans$.

For a set $\markings$ of markings we let $\pre(\markings)$ denote the
set $\setcomp{\marking}{\exists\marking'\in\markingset.\;
\marking\trans\marking'}$, i.e., $\pre(\markings)$ is the set of
markings from which we can reach a marking in $\markingset$ through
the application of a single (timed or discrete) transition.

A set $\markings$ of markings is said to be {\it upward closed} if it
is the case that $\marking\in\markings$ and $\marking\leq\marking'$
imply $\marking'\in\markings$.

\paragraph*{Coverability}
The {\it coverability problem} is defined as follows:
Given a TPN $\tpn$,  a marking $\initmarking$ of $\tpn$, and an upward
closed set $\finalmarkings$ of markings of $\tpn$, is there an
$\marking\in\finalmarkings$ such that $\initmarking\transs\marking$?

Using standard techniques \cite{VW:modelchecking,GoWo:safety}, we can
show that checking several classes of safety properties for TPNs can
be reduced to the coverability problem.
In the next section, we define constraints called {\it existential zones}.
In our reachability algorithm, we use an existential zone to
characterize the set $\finalmarkings$.

\paragraph*{Example}
\begin{figure}
\begin{center}
\epsfig{figure=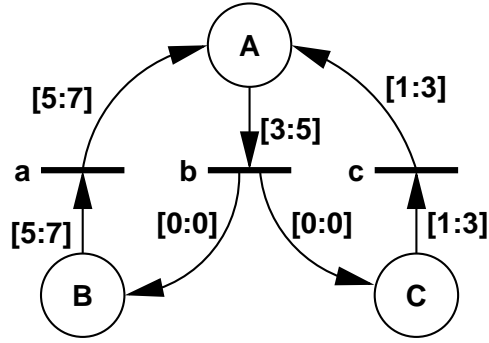, height=4.5cm}
\caption{A small timed Petri net.}
\label{fig:small:example}
\end{center}
\end{figure}
Figure~\ref{fig:small:example} shows an example of a TPN where
$\places=\set{A,B,C}$ and $\transitions=\set{a,b,c}$. For instance,
$\inputs(a)=\tuple{\tuple{B,\intrvl 5 7}}$ and
$\outputs(b)=\tuple{\tuple{B,\intrvl 0 0},\tuple{C,\intrvl 0 0}}$.
The initial marking of this net is the marking
$\initmarking=\tuple{\tuple{A,0.0}}$ with only one token with age
0 in place $A$.

\paragraph*{Remark 1}
For simplicity of presentation we use only non-strict inequalities.
All the results can be generalized in a straightforward manner to
include the more general case, where we also allow strict
inequalities.

\paragraph*{Remark 2}
Notice that, in our definition of the operational behaviour of TPNs,
we assume a lazy (non-urgent) behaviour of the net.
This means that we may choose to ``let time pass'' instead of firing
enabled transitions, even if that makes transitions disabled due to
some of the needed tokens becoming ``too old''. Tokens that are too
old to participate in firing transitions are usually called dead
tokens.
In an urgent TPN, timed transitions that cause dead tokens are not
allowed. This means that the set of transitions of an urgent TPN is a
subset of the set of transitions of the corresponding lazy
TPN. Therefore, if a set of markings is not reachable in the lazy
TPN it is not reachable in the urgent TPN either. In other words safety
properties that hold for the lazy TPN also hold for the urgent TPN. 

\section{Existential Zones}
\label{zones:section}
In this section we introduce a constraint system 
called {\em existential zones}.
Intuitively, an existential zone characterizes an upward closed set of
markings.
An existential zone $\zone$ represents minimal conditions on
markings.
More precisely, 
$\zone$ specifies a minimum number of tokens
which should be in the marking, and then imposes certain
conditions on these tokens.
The conditions are formulated as specifications of the places in which
the tokens should reside and restrictions on their ages.
The age restrictions are stated as bounds on values of clocks, and
bounds on differences between values of pairs of clocks.
A marking $\marking$ which satisfies $\zone$ should have at least the
number of tokens specified by $\zone$.
Furthermore, the places and ages of these tokens should satisfy the
conditions imposed by $\zone$.
In such a case, $\marking$ may have any number of additional tokens
(whose places and ages are irrelevant for the satisfiability 
of the zone by the marking).

For a natural number $n$, we let $\uptoz{n}$ denote the set
$\set{0,1,2,\ldots,n}$, and let $\upto{n}$ denote the set
$\set{1,2,\ldots,n}$.
We assume a TPN $\tpntuple$.

An {\em existential zone $\zone$} is a triple
$\tuple{m,\placing,\diff}$, where $m$ is an natural number, $\placing$
(called a {\it placing}) is a mapping
$\funtype{\placing}{\upto{m}}{\places}$,  and $\diff$ (called a {\em
difference bound matrix}) is a mapping
$\funtype{\diff}{\uptoz{m}\times\uptoz{m}}{\nat\cup\set{\infty}}$.
Intuitively, $m$ defines the minimum number of tokens in the marking,
$\placing$ maps each token to a place, and $\diff$ defines
restrictions on the ages of the tokens in forms of bounds on clock
values and on differences between clock values. Difference bound
matrices, or DBMs, are widely used in verification of timed automata,
e.g., \cite{Dill:DBM,lpw:fct95}.

\begin{figure}
\begin{center}
\epsfig{figure=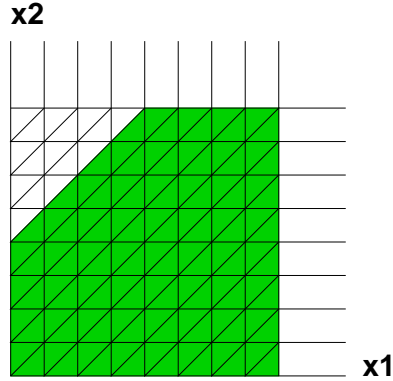, height=5cm}
\caption{Example of restrictions on ages of tokens.}
\label{fig:zone}
\end{center}
\end{figure}
Consider the example from Section~\ref{tpn:section}. Assume that we are
interested in checking the coverability of markings with at least two
tokens, one in place $B$ and one in place $C$, such that the ages of
the tokens are at most 8 and the token in $B$ is at most 4 time units
older than the one in $C$. The markings satisfying these constraints
can be described by the existential zone
$\zone=\tuple{2,\placing,\diff}$ where $\placing(1)=B$,
$\placing(2)=C$ and $\diff$ is described by the following table where
e.g. $\diff(0,i)=0$ and $\diff(2,1)=4$.
\[\diff = \begin{array}{c|ccc}&0&1&2\\\hline 0&-&0&0\\1&8&-&8\\2&8&4&-\end{array}\]
Figure~\ref{fig:zone} shows an illustration of the age restrictions of $\zone$.

Consider a marking
$\marking=\tuple{\tuple{\place_1,x_1},\ldots,\tuple{\place_n,x_n}}$
and an injection
$\funtype{h}{\upto{m}}{\upto{n}}$ (called a {\it witness}).
We say that $\marking$ {\it satisfies $\zone$ with respect to $h$},
written $\marking,h\models\zone$, if the following conditions are
satisfied.
\begin{itemize}
\item
$\placing(i)=\place_{h(i)}$, for each $i:1\leq i\leq m$.
\item
$x_{h(j)}-x_{h(i)}\leq\diff(j,i)$, for each $i,j\in\upto{m}$ with
$i\neq j$.
\item
$x_{h(i)}\leq \diff(i,0)$ and $-\diff(0,i)\leq x_{h(i)}$, for each
$i\in\upto{m}$.
\end{itemize}
We say that $\marking$ {\em satisfies} $\zone$, written
$\marking\models\zone$, if $\marking,h\models\zone$ for some $h$.
Notice that if $\marking$ satisfies $\zone$ then $m\leq n$ (since $h$
is an injection), i.e., $\marking$ has at least the number of tokens
required by $\zone$, and furthermore, the places and ages of the
tokens satisfy the requirements of $\zone$.
We define $\denotationof{\zone}=\setcomp{\marking}{\marking\models\zone}$.
Notice that the value of $\diff(i,i)$ is irrelevant for the
satisfiability of $\zone$.

\paragraph*{Membership}
From the above definitions the following lemma is straightforward.

\begin{lemma}
\label{membership:lemma}
For an existential zone $\zone$ and a marking $\marking$, it is decidable
whether $\marking\models\zone$.
\end{lemma}

\paragraph*{Upward Closedness}
We observe that $\zone$ defines a number of minimal requirements on
$\marking$, in the sense that $\marking$ should contain at least $m$
tokens whose places and ages are constrained by the functions
$\placing$ and $\diff$ respectively.
This means the set $\denotationof{\zone}$ is upward closed since
$\marking\models\zone$ and $\marking\leq\marking'$ implies
$\marking'\models\zone$.

\paragraph*{Normal and Consistent Existential Zones}
An existential zone $\zone=\tuple{m,\placing,\diff}$ is said to be
{\em normal} if for each $i,j,k\in\uptoz{m}$, we have 
$\diff(j,i)\leq\diff(j,k)+\diff(k,i)$.
It is easy to show the following.
\begin{lemma}
\label{zone:normal:lemma}
For each existential zone $\zone$ there is a unique (up to renaming of
the index set) normal existential zone, written $\normal{\zone}$, such
that $\denotationof{\normal{\zone}}=\denotationof{\zone}$.
\end{lemma}
This means that we can  assume without loss of generality that
all existential zones we work with are normal.

An existential zone $\zone$ is said to be {\em consistent} if
$\denotationof{\zone}\neq\emptyset$.

\subsection{Computing Entailment}
We reduce checking entailment between existential zones 
into validity of formulas in
a logic which we here call {\it Difference Bound Logic (DBL)}.
The atomic formulas are either of the form
$\var \leq c$ or of the form 
$\var-\uvar \leq c$, where $\var$ and $\uvar$ are variables
interpreted over $\preals$ and $c \in\nat$.
Furthermore the set of formulas is closed under the propositional connectives.
It is easy to see that validity of DBL-formulas is NP-complete.

Suppose that we are given two existential zones
$\zone_1=\tuple{m,\placing_1,\diff_1}$ and
$\zone_2=\tuple{n,\placing_2,\diff_2}$.
We translate the relation $\zone_1\entailed\zone_2$ into validity
of a DBL-formula $F$ as follows.
We define the set of free variables in $F$ to be
$\setcomp{\var_i}{i\in\upto{n}}$.
Let $H$ be the set of injections from $\upto{m}$ to
$\upto{n}$ such that $h\in H$ if and only if
$\placing_1(i)=\placing_2(h(i))$ 
for each $i\in\upto{m}$.
We define
$F=\left(F_1 \implies \left(\bigvee_{h\in H} F_2\right)\right)$,
where
$F_1=F_{11} \wedge F_{12} \wedge F_{13}$, and 
$F_2=F_{21} \wedge F_{22} \wedge F_{23}$, and

\begin{itemize}
\item
$F_{11}=\bigwedge_{i,j\in\upto{n},j\neq i}
\left(\var_j-\var_i\leq\diff_2(j,i)\right)$.
\item
$F_{12}=\bigwedge_{i\in\upto{n}}
\left(\var_i\leq\diff_2(i,0)\right)$.
\item
$F_{13}=\bigwedge_{i\in\upto{n}}
\left(-\diff_2(0,i)\leq \var_i\right)$.
\item
$F_{21}=\bigwedge_{i,j\in\upto{m},j\neq i}
\left(\var_{h(j)}-\var_{h(i)}\leq\diff_1(h(j),h(i))\right)$.
\item
$F_{22}=\bigwedge_{i\in\upto{m}}
\left(\var_{h(i)}\leq\diff_1(h(i),0)\right)$.
\item
$F_{23}=\bigwedge_{i\in\upto{m}}
\left(-\diff_1(0,h(i))\leq \var_{h(i)}\right)$.
\end{itemize}

This gives the following.

\begin{lemma}
\label{entailment:lemma}
The entailment relation is decidable for existential zones.
\end{lemma}
Notice that in contrast to zones for which entailment can be checked
in polynomial time, the entailment relation for existential zones can
be checked only in nondeterministic polynomial time (as we have to
consider exponentially many witnesses). This is the price we pay for
working with an unbounded number of clocks. On the other hand, when
using zones, the size of the problem grows exponentially with the
number of clocks inside the system.

\subsection{Computing Predecessors}

We define a function $\pre$ such that for a zone $\zone$,
the value of $\pre(\zone)$ is a finite set
$\set{\zone_1,\ldots,\zone_m}$ of zones.
The set  $\pre(\zone)$ characterizes the set of markings
from which we can reach a marking satisfying $\zone$ through the
performance of a single discrete or timed transition.
In other words 
$\pre{\denotationof{\zone}}=\denotationof{\zone_1}\cup\cdots\cup\denotationof{\zone_m}$.
We define $\pre=\pred\cup\preT$, where $\pred$ corresponds to firing
transitions backwards and $\preT$ corresponds to running time backwards.

We define $\pred=\cup_{\transition\in\transitions}\pre_{\transition}$,
where $\pre_{\transition}$ characterizes the effect of running the
transition $\transition$ backwards.
To define $\pre_{\transition}$,
we need the following operations on zones.
In the rest of the section we assume a normal existential zone
$\zone=\tuple{m,\placing,\diff}$, and a timed Petri net
$\tpn=\tpntuple$.
From Lemma~\ref{zone:normal:lemma} we know that assuming $\zone$ to be normal
does not affect the generality of our results.

For an interval $\interval=\intrvl a b$, and $i\in\upto{m}$, we define
the {\em conjunction} $\conjunct{\zone}{\interval}{i}$ of
$\zone$ with $\interval$ at $i$ to be the existential zone
$\zone'=\tuple{m,\placing,\diff'}$, 
where 
\begin{itemize}
\item
$\diff'(i,0)=\min(b,\diff(i,0))$.
\item
$\diff'(0,i)=\min(-a,\diff(0,i))$.
\item
$\diff'(k,j)=\diff(k,j)$,
for each $j,k\in\upto{m}$ with $k\neq j$,  $(k,j)\neq(i,0)$, and
$(k,j)\neq(0,i)$.
\end{itemize}
Intuitively, the operation adds an additional constraint on the age
of token $i$, namely that its age should be in the interval
$\interval$. For example, for a zone 
\[\zone = \tuple{2, \placing, \begin{array}{c|ccc}&0&1&2\\ \hline
0&-&0&0\\ 1&8&-&8\\ 2&8&4&-\end{array}}\] the conjunction
$\conjunct{\zone}{\intrvl 1 6}{1}$ is the zone \[\tuple{2,\placing,\begin{array}{c|ccc}&0&1&2\\\hline 0&-&-1&0\\1&6&-&8\\2&8&4&-\end{array}}\] while the conjunction  $\conjunct{\zone}{\intrvl 0 {10}}{1} = \zone$.

For a place $\place$ and an interval $\interval=\intrvl a b$,
we define the {\em addition} $\add{\zone}{\place}{\interval}$ 
of $\tuple{\place,\interval}$ to $\zone$ to be the existential zone
$\zone'=\tuple{m+1,\placing',\diff'}$, and
\begin{itemize}
\item
$\diff'(m+1,0)=b$, and
$\diff'(0,m+1)=-a$.
\item
$\diff'(m+1,j)=\infty $, and
$\diff'(j,m+1)=\infty$, for each $j\in\upto{m}$.
\item
$\placing'(m+1)=\place$.
\item
$\diff'(k,j)=\diff(k,j)$, for each $j,k\in\uptoz{m}$, and
$\placing'(j)=\placing(j)$, for each $j\in\upto{m}$.
\end{itemize}
Intuitively, the new existential zone $\zone'$ requires one additional 
token to be present in place $\place$ such that
the age of the token is in the interval $\interval$.
For example, for a zone 
\[\zone = \tuple{2,
\begin{array}{c}\placing(1)=B\\\placing(2)=C\end{array},
\begin{array}{c|ccc}&0&1&2\\\hline
0&-&0&0\\1&8&-&8\\2&8&4&-\end{array}}\] the addition
$\add{\zone}{A}{\intrvl 1 2}$ is the zone
\[\tuple{3,\begin{array}{c}\placing(1)=B\\\placing(2)=C\\\placing(3)=A\end{array},
\begin{array}{c|cccc}&0&1&2&3\\\hline 0&-&0&0&-1\\ 1&8&-&8&\infty\\
2&8&4&-&\infty\\ 3&2&\infty&\infty&-\end{array}}\]

For $i\in\upto{m}$, we define the {\em abstraction}
$\abstr{\zone}{i}$ of $i$ in $\zone$
to be the zone $\zone'=\tuple{m-1,\placing',\diff'}$,
where
\begin{itemize}
\item
$\diff'(j,k)=\diff(j,k)$, for each $j,k\in\uptoz{(i-1)}$.
\item
$\diff'(j,k)=\diff(j,k+1)$ and
$\diff'(k,j)=\diff(k+1,j)$, 
for each $j\in\uptoz{(i-1)}$ and $k\in\set{i,\ldots,m-1}$.
\item
$\diff'(j,k)=\diff(j+1,k+1)$, for each $j,k\in\set{i,\ldots,m-1}$.
\item
$\placing'(j)=\placing(j)$, for each $j\in\uptoz{(i-1)}$, and
$\placing'(j)=\placing(j+1)$, for $j\in\set{i,\ldots,m-1}$.
\end{itemize}
Intuitively, the operation removes all constraints related
to token $i$ from $\zone$, so the number of
required tokens is reduced by $1$ and the restrictions related to the
age and place of the token disappear.
For example, for a zone 
\[\zone = \tuple{3,\begin{array}{c}\placing(1)=B\\\placing(2)=C\\\placing(3)=A\end{array},
\begin{array}{c|cccc}&0&1&2&3\\\hline 0&-&0&0&-1\\ 1&8&-&6&7\\
2&8&4&-&7\\ 3&2&2&2&-\end{array}}\]
the abstraction $\abstr\zone 2$ is the zone \[\tuple{2,
\begin{array}{c}\placing(1)=B\\\placing(2)=A\end{array},
\begin{array}{c|ccc}&0&1&2\\\hline 0&-&0&-1\\1&8&-&7\\2&2&2&-\end{array}}\]

Notice that the existential zones we obtain as a result of performing
the three operations above need not be normal.

Now, we are ready to define $\pre$.

\begin{lemma}
\label{pred:lemma}
Consider a TPN $\tpn=\tpntuple$, a transition $\transition\in\transitions$,
and an existential zone $\zone=\tuple{m,\placing,\diff}$.
Let $\inputs(\transition)=
\tuple{\tuple{\place_1,\interval_1},\ldots,\tuple{\place_k,\interval_k}}$,
and $\outputs(\transition)=
\tuple{\tuple{\qplace_1,\jinterval_1},\ldots,\tuple{\qplace_{\ell},\jinterval_{\ell}}}$.
Then $\pret(\zone)$ is the smallest set containing each
existential zone $\zone'$ such that there is a partial injection
$h:\upto{m}\longrightarrow\upto{\ell}$ 
with a domain $\set{i_1,\ldots,i_n}$,
and an existential zone $\zone_1$
satisfying the following conditions.
\begin{itemize}
\item
$\placing(i_j)=\qplace_{h(i_j)}$, for each $j\in\upto{n}$
\item
$\zone\otimes\tuple{\jinterval_{h(i_1)},i_1}\otimes\cdots\otimes
\tuple{\jinterval_{h(i_n)},i_n}$ is consistent.
\item
$\zone_1=\zone\backslash i_1\backslash\cdots\backslash i_n$.
\item
$\zone'=\zone_1\oplus\tuple{\place_1,\interval_1}\oplus\cdots\oplus
\tuple{\place_k,\interval_k}$.
\end{itemize}
\end{lemma}

\begin{lemma}
\label{preT:lemma}
For an existential zone 
$\zone=\tuple{m,\placing,\diff}$,
the set $\preT(\zone)$ is the existential zone 
$\zone'=\tuple{m,\placing,\diff'}$,where
$\diff'(0,i)=0$ and
$\diff'(j,i)=\diff(j,i)$ if
$j\neq 0$, for each $i,j\in\uptoz{m}$, with $i\neq j$.
\end{lemma}

From Lemma~\ref{pred:lemma} and Lemma~\ref{preT:lemma} we get
the following.
\begin{lemma}
\label{pre:lemma}
For an existential zone $\zone$, the set $\pre(\zone)$ is computable.
\end{lemma}

\subsection{BQO}

In order to prove that existential zones are bqo we recall a constraint
system related to existential zones, namely that of {\em existential
regions} introduced in \cite{Parosh:Bengt:Timed:Networks}. 
Let $\cmax$ be the largest natural  number which appears in the
intervals of the given TPN (excluding $\infty$).
An
existential region is a list of multisets $(\mset_0, \mset_1, \ldots, \mset_n,
\mset_{n+1})$ where $n\geq 0$ and $\mset_i$ is a multiset over
$\places\times\uptoz{\cmax}$. In a similar manner to existential zones, an
existential region $\region$ defines a set of conditions which should
be satisfied by a configuration $\conf$ in order for $\conf$ to
satisfy $\region$. Intuitively $\mset_0$ represents tokens with ages
which have fractional parts equal to 0. The multisets $\mset_1, \ldots,
\mset_n$ represent tokens whose ages have increasing fractional parts
where ages of tokens belonging to the same multiset have the same
fractional part and ages of tokens belonging to $\mset_i$ have a
fractional part that is strictly less than the fractional part of the
ages of those in $\mset_{i+1}$. Finally the multiset $\mset_{n+1}$ represents
tokens with ages greater than $\cmax$ (regardless of their fractional
parts).

BQO of existential zones follows from the following two arguments:
\begin{enumerate}
\item
Existential regions are built starting from finite domains, and
repeatedly building finite words, multisets, and sets. From the
properties mentioned above, it follows that existential regions are
bqo.
\item
For each existential zone $\zone$, there is a finite set $\regions$ of
existential regions such that $\zone\equiv\bigvee\regions$. Since bqo is
closed under union, it follows that existential zones are bqo.
\end{enumerate}
This implies the following:
\begin{lemma}
\label{lemma:zone:bqo}
Existential zones are bqo (and hence wqo).
\end{lemma}

\section{Experimental Results}
\label{implementation:section}
We have implemented a prototype to perform coverability analysis for
timed Petri nets. In our experimentation we use a constraint system
called existential DDDs, which is described below. The implementation
is based on a DDD package developed at Technical University of
Denmark~\cite{DDD:impl}. We have used the tool to verify a
parameterized version of Fischer's
protocol \cite{ScBlMa:outlines}.

\subsection{Existential CDDs and DDDs}
{\em Clock Difference Diagrams (CDDs)} \cite{CDD:basic} and {\em
  Difference Decision Diagrams (DDDs)} \cite{DDD:basic} are constraint
systems that have been invented to give representations of real-time
systems that are more compact than zones. In the same manner as zones
were modified into existential zones, we modify the definitions of
CDDs (DDDs) into {\em existential CDDs (DDDs)} \index{existential CDD}
\index{existential DDD} to make them suitable
for verifying systems with an unbounded number of clocks. Below we
give the definition of existential DDDs. The definition of existential
CDDs can be stated in a similar manner.

An {\em existential DDD} $\ddd$ is a tuple
$\tuple{m,\placing,\vertices,\edges}$, where $m$ is a natural
number denoting the minimum number of tokens in a marking satisfying
$\ddd$ and the placing $\placing$ maps each token to a place in the
same manner as in an existential zone
(Section~\ref{zones:section}). $\tuple{\vertices,\edges}$ is a finite
directed acyclic graph where $\vertices$ is the set of vertices and
$\edges$ is the set of edges. The set $\vertices$ contains two
special elements $\refvertex$ and $\reffvertex$. The out-degrees of
$\refvertex$ and $\reffvertex$ are zero while the out-degrees of the
rest of vertices are two. Each vertex
$\vertex\in\vertices\setminus\set{\refvertex,\reffvertex}$ has the
following attributes: $\positive(\vertex),\negative(\vertex)\in\uptoz
m$, $\op(\vertex)\in\set{<,\leq}$, $\const(\vertex)\in\integers$, and
$\high(\vertex),\low(\vertex)\in\vertices$. The set $\edges$ contains
the edges $\tuple{\vertex,\low(\vertex)}$ and
$\tuple{\vertex,\high(\vertex)}$, where
$\vertex\in\vertices-\set{\refvertex,\reffvertex}$. In a similar
manner to BDDs, the internal nodes of $\ddd$ correspond to the
if-then-else operator $\phi\rightarrow\phi_1,\phi_2$, defined as
$(\phi\wedge\phi_1) \vee (\neg\phi\wedge\phi_2)$. Intuitively, the
attributes of the node represent the DBL-formula
$\phi={x_{\positive(\vertex)}-x_{\negative(\vertex)}}\,\op(\vertex)\,\const(\vertex)$,
and $\high(\vertex)$ and $\low(\vertex)$ are children of $\vertex$
corresponding to $\phi_1$ and $\phi_2$ respectively. The special
vertices $\refvertex$ and $\reffvertex$ correspond to {\em false} and
{\em true}.

Consider an existential DDD
$\ddd=\tuple{m,\placing,\vertices,\edges}$, a vertex
$\vertex\in\vertices$, a marking
$\marking=\tuple{\tuple{\place_1,x_1}, \ldots, \tuple{\place_n,x_n}}$
and an injection $\funtype{h}{\upto m}{\upto{n}}$. We say that {\em
  $\marking$ satisfies $\ddd$ at $\vertex$ with respect to $h$},
written $\marking,h\models\tuple{\ddd,\vertex}$, if
$\placing(i)=\place_{h(i)}$, for each $i\in\upto m$, and either
\begin{itemize}
\item
$\vertex=\reffvertex$; or
\item
$\left(\left(
\begin{array}{c}
x_{h(\positive(\vertex))}\\
-\\
x_{h(\negative(\vertex))}
\end{array}
\right)
\sim\const(\vertex)
\right)
\rightarrow 
\left(
\begin{array}{c}
\marking,h\models\tuple{\ddd,\high(\vertex)}\\
,\\
\marking,h\models\tuple{\ddd,\low(\vertex)}
\end{array}
\right)
$\hfill\\[.5em]
where $\sim=\op(\vertex)$.
\end{itemize}
As with existential zones, we can modify the operations defined in
\cite{DDD:basic} to compute predecessors of existential DDDs with
respect to transitions of a timed Petri net. To check entailment we
must, as we did for existential zones, take into consideration all
variable permutations.

For each existential DDD $\ddd$ there is a finite set $\zones$ of
existential zones such that
$\denotationof{\ddd}=\denotationof{\bigvee\zones}$. Intuitively this means
that an existential DDD can replace several existential zones, and
hence existential DDDs give a more compact (efficient) representation
of sets of states. 
From Lemma~\ref{lemma:zone:bqo},
Theorem~\ref{bqo:properties:theorem} (Property~\ref{bqo:disj:prop}),
and the fact that 
each existential DDD is the disjunction of a finite set of
existential zones we get the following result.

\begin{lemma}
\label{lemma:DDD:bqo}
Existential DDDs are better quasi-ordered (and hence also well quasi-ordered).
\end{lemma}

\subsection{Fischer's Protocol}
We will now describe a timed Petri net model of a parameterized
version \index{parameterized system} of Fischer's
protocol~\cite{ScBlMa:outlines}. The purpose of the protocol is to
guarantee mutual exclusion in a concurrent system consisting of an
arbitrary number of processes. The example was suggested by Schneider et
al.~\cite{ScBlMa:outlines}. The protocol analyzed here is in fact a
weakened version of Fischer's protocol but since the set of reachable
states of the weakened version is a superset of the reachable states
of the original protocol, the results of our analysis are still valid.

\begin{figure}
\begin{center}
\includegraphics[width=.8\textwidth]{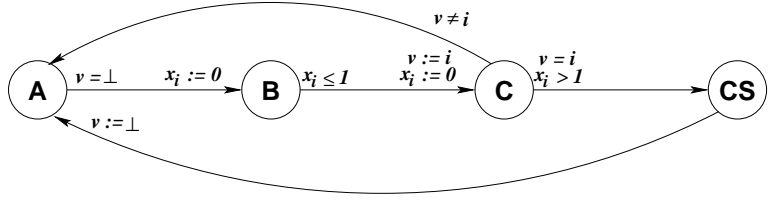}
\caption{Fischer's Protocol for Mutual Exclusion}
\label{fig:fischer_simple}
\end{center}
\end{figure}
The protocol consists of each process running the code that is
graphically described in Figure~\ref{fig:fischer_simple}. Each process
$i$ has a local clock, $x_i$, and a control state that assumes
values in the set $\set{A,B,C,C\!S}$ where $A$ is the initial state
and $C\!S$ is the critical section. The processes read from and write
to a shared variable $v$, whose value is either $\bot$ or the index
of one of the processes.

All processes start in state $A$. If the value of the shared variable
is $\bot$, a process wishing to enter the critical section can
proceed to state $B$ and reset its local clock. From state $B$, the
process can proceed to state $C$ within one time unit or get stuck in
$B$ forever. When making the transition from $B$ to $C$, the process
resets its local clock and sets the value of the shared variable to 
its own index. The process now has to wait in state $C$ for more than
one time unit, a period of time that is strictly greater than the one
used in the timeout of state $B$. If the value of the shared variable
is still the index of the process, the process may enter the critical
section, otherwise it may return to state $A$ and start over
again. When exiting the critical section, the process resets the
shared variable to $\bot$.

\begin{figure}
\begin{center}
\includegraphics[width=\textwidth]{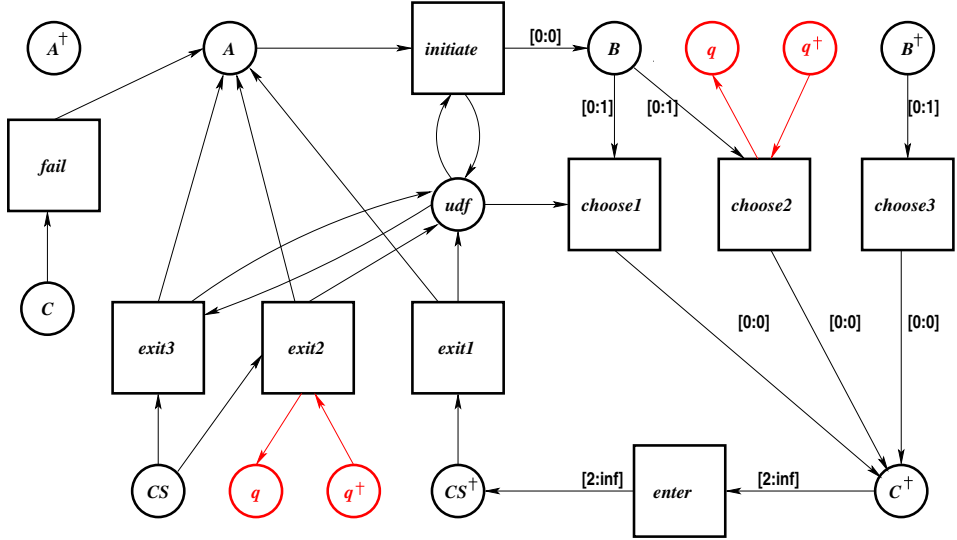}
\caption{Timed Petri net model of Fischer's Protocol for Mutual Exclusion}
\label{fig:fischer_tpn}
\end{center}
\end{figure}

We will now make a model of the protocol in our timed Petri net formalism.
The processes running the protocol are modelled by tokens in the places
$A$, $B$, $C$, $C\!S$, $A^{\dagger}$, $B^{\dagger}$, $C^{\dagger}$ and
$C\!S^{\dagger}$. The places marked with $\dagger$ represent that the
value of the shared variable is the index of the process modelled by
the token in that place. We use a place $\mathit{udf}$ to represent that the
value of the shared variable is $\bot$. A straightforward translation
of the description in Figure~\ref{fig:fischer_simple} yields the 
Petri net model in Figure~\ref{fig:fischer_tpn}.
$q$ is used to denote an arbitrary process state. 
The critical section is modelled by the places $C\!S$ and $C\!S^{\dagger}$, so
mutual exclusion is satisfied when the number of tokens in those places
is less than two.

\subsection{Results}
We have used our prototype to analyze the parameterized version of
Fischer's protocol presented above.
In order to prove mutual exclusion we examine the reachability of the
existential zones stating that at least two processes are in the
critical section, i.e., the following zones:
\begin{itemize}
\item $\zone_1=\tuple{2,\placing_1,\diff}$ where $\placing_1(1)=\placing_1(2)=C\!S$
\item $\zone_2=\tuple{2,\placing_2,\diff}$ where $\placing_2(1)=C\!S$ and $\placing_2(2)=C\!S^{\dagger}$
\item $\zone_3=\tuple{2,\placing_3,\diff}$ where $\placing_3(1)=\placing_3(2)=C\!S^{\dagger}$
\end{itemize}
For all three zones $\diff(0,i)=0$, $\diff(i,j)=\infty$ for $i\neq j$.

The reachable state space,
represented by 45 existential DDDs, takes 3.5 seconds to compute on a
Sun Ultra 60 with 512 MB memory and a 360 MHz UltraSPARC-II
processor. In the process, $\pre$ was computed for 51 existential DDDs.

\section{Broadcast Protocols}
\label{broadcast:section}
We consider {\it broadcast protocols}, which consist of an arbitrary number
of identical finite-state processes, communicating through rendezvous
or through broadcast.
We assume a finite set $\set{s_1,\ldots,s_n}$ of {\em states}, and a set
$\set{x_1,\ldots,x_n}$ of variables which range over the natural numbers.
A {\em configuration $\conf$} of a protocol is a tuple
$\tuple{a_1,\ldots,a_n}$ of natural numbers, where $a_i$
represents the number of processes which are in the state $s_i$.
In \cite{Esparza:Finkel:Mayr:LICS} a constraint system, which we here
refer to as $\broadconstrs$, is defined, where each constraint
is a tuple $\tuple{b_1,\ldots,b_n}$ with
a denotation $\denotationof{\tuple{b_1,\ldots,b_n}}$ which is the 
upward closed set 
$\setcomp{\tuple{a_1,\ldots,a_n}}{\tuple{b_1,\ldots,b_n}\leq\tuple{a_1,\ldots,a_n}}$.
In \cite{Delzanno:etal:broadcast:constr} several new constraint systems for
broadcast protocols are proposed, 
and compared with regard to the efficiency parameters mentioned
in Section~\ref{bqo:application:section}.
The most general of these constraint systems , called {\bf AD} in
\cite{Delzanno:etal:broadcast:constr},
consists of conjunctions of constraints each of the form
$x_{i_1}+\cdots+x_{i_k}\geq b$, where 
$x_{i_1},\ldots,x_{i_k}$ are distinct variables of $\set{x_1,\ldots,x_n}$.
Two special cases are considered:
{\bf NA} where $k$ is always equal to $1$, and {\bf DV} where
the set of variables occurring in the different conjuncts are assumed
to be disjoint.
Since these new constraint systems are not constructed applying the basic set
of constraint operations (described in Section~\ref{bqo:application:section}),
a separate proof of termination is required for them.

Applying the method of Section~\ref{bqo:application:section} we can show
bqo of {\bf AD}, {\bf NA}, and {\bf DV} uniformly as follows.
From properties \ref{finite:bqo:prop} and \ref{bqo:str:prop} in
Theorem~\ref{bqo:properties:theorem}, it follows that
$\broadconstrs$ is bqo.
Furthermore, it is straightforward to show that each constraint in
{\bf AD}, {\bf NA}, and {\bf DV}  is equivalent to the disjunction of
a finite set of constraints in $\broadconstrs$.
From property~\ref{bqo:disj:prop} 
of Theorem~\ref{bqo:properties:theorem}, we get
\begin{theorem}
\label{broad:bqo:theorem}
{\bf AD}, {\bf NA}, and {\bf DV} are bqo.
\end{theorem}
In fact we can derive the bqo property for a more general constraint
system than {\bf AD}, namely that consisting of basic constraints of the form
$a_1 x_1+\cdots+a_k x_k\geq b$, combined through conjunction and disjunction.

\section{Lossy Channel Systems}
\label{lcs:section}
In \cite{AbJo:lossy:IC}, we present a constraint system, 
here denoted $\lcsconstrs_1$, for
representing upward closed sets of words.
The constraints in $\lcsconstrs_1$ are used in \cite{AbJo:lossy:IC}
for verification
of {\it lossy channel systems}: finite state machines communicating
over unbounded and unreliable FIFO buffers.
We assume a finite alphabet $\Sigma$.
For words $w_1,w_2\in\Sigma^*$, we let
$w_1\wpreceq w_2$ denote that $w_1$ is 
a (not necessarily contiguous) subword of $w_2$.
A constraint in $\lcsconstrs_1$ is represented by a word
$w$, where  $\denotationof{w}=\setcomp{w'}{w\wpreceq w'}$.

Here, we introduce a new constraint system $\lcsconstrs_2$, defined as
the smallest set such that $\lcsconstrs_2$ contains:
\begin{itemize}
\item
$a$, for each $a\in\Sigma$, where $\denotationof{a}=\setcomp{w}{a\wpreceq w}$;
\end{itemize}
and $\lcsconstrs_2$ is closed under:
\begin{itemize}
\item
concatenation:

$\denotationof{\constr_1\bullet\constr_2}=
\setcomp{w_1w_2}{w_1\in\denotationof{\constr_1}\;\mbox{and}\;
w_2\in\denotationof{\constr_2}}$;
\item
conjunction:

$\denotationof{\constr_1\&\constr_2}=
\setcomp{w}{w\in\denotationof{\constr_1}\;\mbox{and}\;
w\in\denotationof{\constr_2}}$; and
\item
disjunction:

$\denotationof{\constr_1+\constr_2}=
\setcomp{w}{w\in\denotationof{\constr_1}\;\mbox{or}\;
w\in\denotationof{\constr_2}}$.
\end{itemize}

\begin{figure}
\begin{center}
\includegraphics[width=8.2cm]{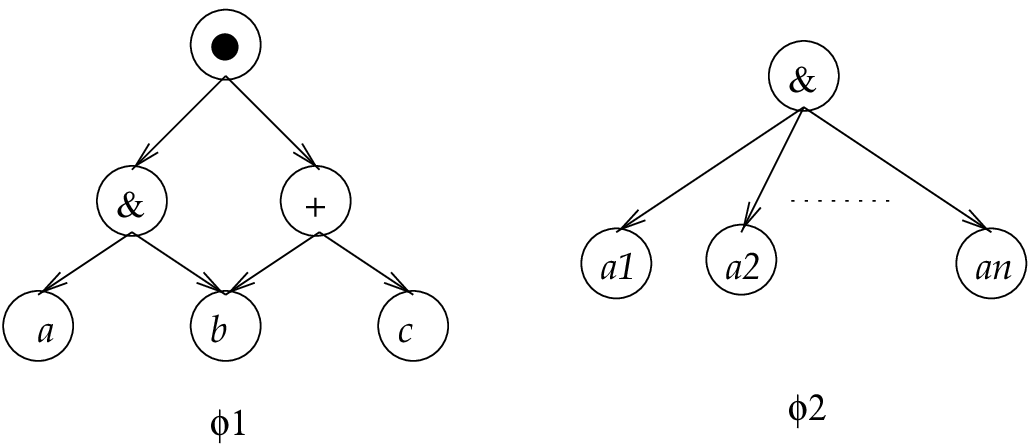}
\caption{Two constraints in $\lcsconstrs_2$}.
\label{lcs:figure}
\end{center}
\end{figure}
\begin{example}
In Figure~\ref{lcs:figure}, the constraint $\phi_1$ is of the form
$(a\;\&\;b)\bullet(b+c)$.
This means that
$\denotationof{\phi_1}=
\setcomp{w_1w_2}{(a\wpreceq w_1)\;\mbox{and}\;(b\wpreceq w_1)\;\mbox{and}\;
((b\wpreceq w_2)\;\mbox{or}\;(c\wpreceq w_1))}$.
The constraint $\phi_1$ is equivalent
to the disjunction of the following set of constraints in
$\lcsconstrs_1$: $\set{abb,abc,bab,bac}$.
\end{example}
The constraint system $\lcsconstrs_2$ is exponentially more succinct than
 $\lcsconstrs_1$.
More precisely,
each constraint $\constr_1\in\lcsconstrs_1$ has a
linear-size translation (through the concatenation operator)
into an equivalent constraint $\constr_2\in\lcsconstrs_2$.
On the other hand a constraint of the form
$a_1\&\cdots\& a_n$ ($\constr_2$ in Figure~\ref{lcs:figure})
can only be represented in $\lcsconstrs_1$
by the disjunction of a set of constraints of size $n!$; namely
the set \\
$\setcomp{b_1\bullet\cdots\bullet b_n}
{\tuple{b_1,\ldots,b_n}\;\mbox{is a permutation of}\; \tuple{a_1,\ldots,a_n}}$.
 
In a similar manner to Section~\ref{zones:section} and
Section~\ref{broadcast:section} we can use properties of
$\lcsconstrs_1$ and $\lcsconstrs_2$ to conclude the following
\begin{theorem}
\label{lcs:bqo:theorem}
$\lcsconstrs_2$ is bqo.
\end{theorem}

\section{Integral Relational Automata}
\label{ira:section}
An {\em Integral Relational Automaton (IRA)} operates on  a set
$X=\set{x_1,\ldots,x_n}$ of {\em variables} 
assuming values from the set $\integers$ of integers.
The transitions of the automaton are labeled
by guarded commands of the form $g \rightarrow stmt$ in which
the guard $g$  is a boolean combination of inequalities of form
$x < y$, $c < x$, or $x < c$, for $x,y\in X$ and $c\in\integers$; and where
the body $stmt$ contains, for each $x \in X$, an assignment
of one of the forms
$x := y$, $x := c$, or $x := \set{?}$, for
$y \in X$ and $c\in\integers$.
The assignment $x := \set{?}$ is a ``read'' operation putting
an arbitrary integer into the variable $x$.
A {\em configuration $\conf$} of an IRA is a mapping
from $X$ to $\integers$.
Sometimes, we write $\conf$ as a tuple
$\tuple{\conf(x_1),\ldots,\conf(x_n)}$.
For $c\in\integers$, we use the convention that $\conf(c)=c$.

A constraint system, called the {\em sparser than} system $\sparseconstrs$,
is defined in \cite{Cerans:relational:automata:ICALP}, for verification of IRAs
as follows.
Let $c_{\it min}$ ($c_{\it max}$) be the smallest (largest) constant
occurring syntactically in the IRA.
Define $C=\set{c_{\it min},\ldots,c_{\it max}}$ to be the set of integers
between $c_{\it min}$ and $c_{\it max}$.
A constraint $\constr$ in $\sparseconstrs$ is a mapping
from $X$ to $\integers$.
In a similar manner to configurations, we assume
$\conf(c)=c$ for $c\in\integers$.
A configuration $\conf$  satisfies $\constr$ iff
for each $x,y\in X\cup C$, we have
(i) $\conf(x)\leq\conf(y)$ iff
$\constr(x)\leq \constr(y)$; and 
(ii) if $\constr(x)\leq \constr(y)$ then
$\constr(y)-\constr(x)\leq\conf(y)-\conf(x)$.
\begin{example}
\label{ira:old:example}
Assume $X=\set{x_1,x_2,x_3}$ and $C=\set{5}$.
Consider a constraint $\constr=\tuple{10,5,12}$, then
$\conf_1=\tuple{12,5,17}\in\denotationof{\constr}$, while
$\conf_2=\tuple{8,5,16}\not\in\denotationof{\constr}$
(since $\constr(x_1)-\constr(x_2)=5\not\leq\conf_2(x_1)-\conf_2(x_2)=3$), and
$\conf_3=\tuple{12,4,17}\not\in\denotationof{\constr}$
(since $\constr(5)=5\leq\constr(x_2)=5$ while $\conf_3(5)=5\not\leq\conf_3(x_2)=4$).
\end{example}

We introduce a new constraint system $\exsparseconstrs$, such that
a constraint $\constr$ in $\exsparseconstrs$ is a
conjunction of conditions of the forms
$c\leq x$, $x\leq c$, and $c\leq y-x$, where $x,y\in X$ and $c\in\integers$.
The satisfiability of $\constr$ by a configuration
$\conf$ is defined in the obvious way.
\begin{example}
Assume $X=\set{x_1,x_2}$ and $C=\set{5}$.
The constraint $5<x_2$ in $\exsparseconstrs$ is equivalent to
the disjunction of the following set of constraints in
$\sparseconstrs$: \\
$\set{\tuple{4,7},\tuple{5,7},\tuple{6,7},\tuple{7,7},\tuple{8,7}}$.
Notice that the constraints correspond to the different relative values which
$x_1$ may have with respect to the constant $5$ and the variable $x_2$.
\end{example}

In a similar manner to the constraint systems
in the previous sections, we can
show that $\exsparseconstrs$ is exponentially more succinct than
$\sparseconstrs$ and that the following theorem holds.
\begin{theorem}
\label{ira:bqo:theorem}
$\exsparseconstrs$ is bqo.
\end{theorem}

\section{Conclusions and Future Work}
\label{conclusion:section}

We have proposed {\em better quasi-orderings}, a refinement of the
theory of {\em well quasi-ordering}, as a framework for symbolic model
checking since they allow us to build constraint systems which are
more compact than previous ones.
For instance, we show better quasi-ordering of complex expressions for
upward closed sets of words, used for verification of lossy channel
systems and for arbitrary boolean combinations of linear inequalities,
used for verification of broadcast protocols.
We also achieve similar results for binary constraints which can be
applied for model checking of real-time systems and relational
automata.

We have introduced a new constraint system, {\em existential zones}
for verification of real-time systems with an unbounded number of
clocks. Using and modifying efficient data structures for verification
of real-time automata, we have obtained some encouraging experimental
results.
One direction for future work is to design efficient data
structures for manipulating the new constraint systems. 
It would also be interesting to investigate the feasibility of
defining a general framework for implementation of better
quasi-ordered constraint systems.

Furthermore, in addition to disjunction, better quasi-orderings are
closed under several other operations which do not preserve well
quasi-ordering. An example is that better quasi-orderings are closed
under the operation of taking infinite sets and infinite words. This
means that we can consider much richer structures for building
constraints.
Therefore, although the main concern of this work is that of efficiency,
we believe that the approach will eventually also lead to decidability
results for new classes of infinite-state systems.

\section*{Acknowledgments}
We are grateful to Alberto Marcone and Petr Jan\v{c}ar for 
many interesting discussions on the theory of better quasi-orderings.
Many thanks to Ahmed Bouajjani, Purushothaman Iyer, 
and Pritha Mahata for comments on earlier versions of the paper.
Special thanks to Jesper M{\o}ller and Henrik Reif Andersen for letting 
us use their DDD implementation.

\bibliographystyle{alpha}
\bibliography{bibdatabase}
\end{document}